\documentclass[12pt]{article}
\usepackage{color}
\newcommand{\beq}{\begin{equation}}
\newcommand{\eeq}{\end{equation}}
\newcommand{\beqa}{\begin{eqnarray}}
\newcommand{\eeqa}{\end{eqnarray}}
\newcommand{\beqar}{\begin{eqnarray*}}
\newcommand{\eeqar}{\end{eqnarray*}}

\newcommand{\al}{\alpha}
\newcommand{\be}{\beta}

\def\non          {\nonumber}
\def\ha           {\mbox{$\frac{1}{2}$}}

\def\Tr           {\mbox{\rm Tr}\,}

\def\cd           {{\cdot}}
\def\ran          {\rangle}
\def\lan          {\langle}
\def\fsk    {k\!\!\!\!/\,}
\def\fsH    {H\!\!\!\!/\,}
\newcommand{\del}{\delta}

\newcommand{\ga}{\gamma}

\newcommand{\lam}{\lambda}

\newcommand\bPsi{{\bar \Psi }}

\newcommand{\z}{\zeta}

\newcommand{\labell}[1]{\label{#1}} 
\newcommand{\reef}[1]{(\ref{#1})}

\newcommand\veps{\varepsilon}

\newcommand\cD{{\cal D}}

\newcommand\bu{\bar{u}}

\def\sst#1{{\scriptscriptstyle #1}}
\def\0{{\sst{(0)}}}
\def\1{{\sst{(1)}}}
\def\2{{\sst{(2)}}}
\def\3{{\sst{(3)}}}
\def\4{{\sst{(4)}}}
\def\5{{\sst{(5)}}}
\def\6{{\sst{(6)}}}
\def\7{{\sst{(7)}}}
\def\8{{\sst{(8)}}}

\begin{document}
\baselineskip 18pt%
\begin{titlepage}
\vspace*{1mm}%
\hfill
\vbox{

    \halign{#\hfil         \cr
           } 
      }  
\vspace*{9mm}
\vspace*{9mm}%

\center{ {\bf \Large  Closed string Ramond-Ramond proposed higher derivative
interactions on fermionic amplitudes in IIB
}}\vspace*{1mm} \centerline{{\Large {\bf  }}}
\begin{center}
{Ehsan Hatefi }

\vspace*{0.8cm}{ {\it
International Centre for Theoretical Physics\\
 Strada Costiera 11, Trieste, Italy  }}
\footnote{ehatefi@ictp.it}

\vspace*{0.1cm}
\vspace*{.1cm}
\end{center}
\begin{center}{\bf Abstract}\end{center}
\begin{quote}
The complete form of the amplitude of one closed string Ramond-Ramond (RR), two fermionic strings and one scalar field in IIB superstring theory has been computed in detail. Deriving  $<V_{C}V_{\bar\psi}V_{\psi} V_{\phi}>$  by  using suitable gauge fixing, we discover some new vertices and their higher derivative corrections. We investigate both infinite gauge and scalar $u-$channel poles of this amplitude. In particular, by using the fact that the kinetic term of fermion fields has no correction, employing  Born-Infeld action, the Wess-Zumino terms and their higher derivative corrections, we discover all  infinite $t,s-$channel fermion poles. The couplings between one RR and two fermions and all their infinite higher derivative corrections have been explored. In order to look for all infinite $(s+t+u)-$ channel scalar/gauge poles for $p+2=n,p=n$ cases, we obtain the couplings between two fermions-two scalars and  two fermions, one scalar and one gauge field as well as all their infinite higher derivative corrections in type IIB. Specifically we make various comments based on arXiv:1205.5079 in favor of  universality conjecture for all order  higher derivative corrections (with or without low energy expansion) and the relation of open/closed string  that is responsible for all superstring scattering amplitudes in IIA,IIB.
\end{quote}
\end{titlepage}

\section{Introduction}

D-branes  \cite{Polchinski:1995mt},\cite{Witten:1995im},\cite{Polchinski:1996na}
 have been clarifying essential tools in most of the progresses in theoretical high energy physics as well as in  superstring theories. As an instance of dynamical aspects of D-branes, one may consider the transition between both open and closed string\cite{Ademollo:1974fc}.  D-brane physics has diverse dual descriptions  (see string dualities \cite{Polchinski:1996nb}).
 Some more examples like the $D0/D4$ system with their explanations have been realised in the introduction of  \cite{Hatefi:2012sy}. In particular in \cite{Hatefi:2012bp} we have shown how the world volume must appear from super gravity point of view, namely we set new kind of ADM reduction to IIB which was reduced to hyperboloidal $H^5$ space in 5D and showed that A(ds) brane world might be understood.

In IIA(IIB) string theories, D$_p$-branes with  even (odd) $p$ (which is the spatial dimension of a  D$_p$-brane) are related to BPS branes in which there is no instability and supersymmetry is not broken. Apart from these properties, BPS branes carry Ramond-Ramond (RR) charge. To describe dynamics of branes, one has to work  with some proper effective actions, namely one has to deal with  bosonic effective actions and also should distinguish it from its supersymmetric  version. Bosonic actions in the presence of various D$_p$-brane configurations  were considered in \cite{Myers:1999ps}.
\vskip 0.1in

The  supersymmetrized version of bosonic effective action that appeared in \cite{Myers:1999ps}  has not been completely found yet, however it is worth mentioning  \cite{Howe:2006rv} as an important reference. In this paper, in order to avoid some details we address some of the main references. One might start reading \cite{Leigh:1989jq} to explore the effective action for a  bosonic D$_p$-brane. To follow  the supersymmetric action for a  D$_p$-brane reference \cite{Cederwall:1996pv} should be considered.

 \vskip .02in
For a comprehensive review of Myers terms, the Chern-Simons action, the Wess-Zumino actions and more significantly for Born-Infeld action \cite{Hatefi:2010ik,Hatefi:2012zh} and all references therein are suggested. The three standard effective field theory methods namely Pull-back, Taylor expansion and Myers terms have been addressed in \cite{Hatefi:2012wj}.  Some methods for looking for all  the higher derivative corrections of Myers, Chern-Simons and Born-Infeld actions, have been expressed in \cite{Hatefi:2012wj}.

\vskip 0.02in

Let us point out an issue in favor of scattering theory in string theory. We may hint to a conjecture which appeared in \cite{Park:2007mc} where BPS open strings quantum effects might indicate the host branes curvature. Given some attempts \cite{Koerber:2002zb,Keurentjes:2004tu,Denef:2000rj}, it would be nice to find the complete form of Wess-Zumino (WZ) and DBI actions. In this paper we provide some more data and our S-matrix will be useful for all order $\alpha'$ determination of DBI and WZ actions. We just refer to \cite{Hashimoto:1996bf} out of so many works that involved either the scattering from stable branes or dealt with intriguing applications of the branes. Note that some of our higher derivative corrections  come from the couplings of lower dimensional branes with closed string RR, meanwhile lower dimensional branes should be realised as some soliton objects. To be more specific, in \cite{Hatefi:2012sy} we explored dissolving lower dimensional branes inside  higher dimension branes \cite{Hatefi:2012wz}.  Another example is related to $D(-1)/D3$ configuration, where this system has $N^2$
entropy behavior and this result can be interpreted by taking into account higher order $\alpha'$ Myers terms. Some applications of new couplings (including their corrections) in M-theory are recently addressed in \cite{McOrist:2012yc,Hatefi:2012wz}.
\vskip 0.1in


Having set some of the past works on Myers terms and WZ effective actions \cite{Hatefi:2010ik},\cite{Hatefi:2012ve,Hatefi:2012rx,Hatefi:2012cp,Garousi:2007fk,Hatefi:2008ab}, we would like to explore  all the infinite  two fermions -two scalars and two fermions, one scalar and one gauge field  couplings as well as  their all order $\alpha'$  higher derivative corrections.  Basically we want to find out all the infinite effective couplings between  two fermions two scalars and two fermions, a scalar and a gauge field by matching field theory vertices with  an infinite number of scalar/gauge $(t+s+u)$-channel poles of the string amplitude of $<V_{C} V_{\phi} V_{\bar\psi}V_{\psi} >$ for $p+2=n$, $p=n$ cases ($n$ is the rank of RR field strength $H$) accordingly.  These new couplings which we intend to derive might have some applications to F-theory  \cite{Vafa:1996xn}  as well as  M-theory \cite{Hatefi:2012sy,Hatefi:2012wz,McOrist:2012yc}.

\vskip 0.1in

 This work illustrates the fact that super Yang-Mills vertices such as two fermion-two scalar couplings and in particular two fermion-one gauge-one scalar couplings give rise precisely to the same scalar/ gauge poles as those poles which appeared in the string theory amplitude of $<V_{C} V_{\phi} V_{\bar\psi}V_{\psi} >$. This paper is organised as follows.

 In the next section, we explain   superstring scattering computations  of a closed string RR field, two fermion fields and one scalar field, to actually obtain the complete and closed form of the correlators of
 $<V_{C} V_{\phi} V_{\bar\psi}V_{\psi} >$. Our computations make sense in IIB superstring theory as both fermions here carry the same chirality.

\vskip 0.06in
 Then by expanding the amplitude at low energy limit  and by finding  the desired vertices such as
 the vertex of one RR, two fermion fields and its extensions (to all orders in $\alpha'$) as well as using WZ terms, we  produce both  infinite scalar and gauge $u$-channel poles for  $p+2=n$ and $p=n$ cases. We move on to produce all infinite $t,s$-channel fermion poles by obtaining the infinite extensions of the vertex of one RR and two fermions. Finally we summarise  our results and talk about  the deep relation that exists between a closed string and an open string  and we find that indeed it is responsible for matching superstring amplitudes with their effective field theories. In order to avoid introducing some more details and notations, we refer the reader to Appendix A and B of \cite{Hatefi:2012wj,Hatefi:2012ve}.
\vskip 0.1in
This paper provides some more information on the universal behavior of the higher derivative corrections \cite{Hatefi:2012rx}. In particular the calculations of this paper clarify that the universal conjecture for higher order corrections which appeared in \cite{Hatefi:2012rx} works even for fermionic amplitudes, including two fermion-two scalar couplings. Thus our S-matrix serves  one more test of our understanding of the full DBI, WZ effective actions. This universal conjecture might also be useful for deriving all the singularities of the higher point functions of BPS  amplitudes without the need for applying direct conformal field theory  computations. This universal behavior should have origins coming from  the deep relation of a closed and an open string.
This relation may clarify closed string's composite nature only in terms of open strings. We describe it further in the conclusion section.
\section{ Notations and Analysis of  $<V_{\bar\psi}V_{\psi}V_{\phi}>$ }

In this section we use conformal field theory methods to actually find out  the entire amplitude of  a closed string RR (C-field), two fermions (with the same chirality) and
one scalar field  in the world volume of IIB superstring theory.

Given the vast recent research works on scattering amplitudes, it is indeed impossible to address all
attempts on  this subject, however we would like to address some of the works  that carried out at tree level computations
\cite{Stieberger:2009hq,Kennedy:1999nn,Chandia:2003sh,Hatefi:2010ik,Hatefi:2012ve,Hatefi:2012rx}.
The needed vertex operators for our purpose are \footnote{We clearly consider  $\alpha'$
 in this paper
 .}
\beqa
V_{\phi}^{(0)}(x) &=& \xi_{i}\bigg(\partial
X^i(x)+\alpha' ik\cd\psi\psi^i(x)\bigg)e^{\alpha' ik\cd X(x)},
\nonumber\\
V_{\bPsi}^{(-1/2)}(x)&=&\bu^Ae^{-\phi(x)/2}S_A(x)\,e^{\alpha'iq.X(x)}, \nonumber\\
V_{\Psi}^{(-1/2)}(x)&=&u^Be^{-\phi(x)/2}S_B(x)\,e^{\alpha'iq.X(x)}, \nonumber\\
V_{C}^{(-\frac{1}{2},-\frac{1}{2})}(z,\bar{z})&=&(P_{-}\fsH_{(n)}M_p)^{\al\be}e^{-\phi(z)/2}
S_{\al}(z)e^{i\frac{\alpha'}{2}p\cd X(z)}e^{-\phi(\bar{z})/2} S_{\be}(\bar{z})
e^{i\frac{\alpha'}{2}p\cd D \cd X(\bar{z})},
\label{d4Vs}
\eeqa
  $(p,q,k)$ become  the momenta
of the RR, fermion and scalar field accordingly. Their  on-shell condition is
$k^2=q^2=p^2=0$.
Note that  $u^A$ is the fermion 's wave function of  Majorana-Weyl  in ten dimensions of space-time. Spin indices have been raised by  $C^{\alpha\be}$  (charge conjugation matrix) ,
\beqa
(P_{-}\fsH_{(n)})^{\al\be} =
C^{\al\del}(P_{-}\fsH_{(n)})_{\del}{}^{\be},
\eeqa
where the definitions of the traces are
\beqa
\Tr(P_{-}\fsH_{(n)}M_p\gamma^{k})&\equiv & (P_{-}\fsH_{(n)}M_p)^{\alpha\beta}(\gamma^{k}C^{-1})_{\alpha\beta}
\nonumber\\
\Tr(P_{-}\fsH_{(n)}M_p\Gamma^{iab})&\equiv &
(P_{-}\fsH_{(n)}M_p)^{\alpha\beta}(\Gamma^{iab}C^{-1})_{\alpha\beta},
\eeqa
where $P_{-}$ is a projection operator, $P_{-} = \ha (1-\ga^{11})$  and the field strength of RR is
\begin{displaymath}
\fsH_{(n)} = \frac{a
_n}{n!}H_{\mu_{1}\ldots\mu_{n}}\ga^{\mu_{1}}\ldots
\ga^{\mu_{n}}
\ ,
\non\end{displaymath}
   where for  type IIB theory,  $n=1,3,5$ and  $a_n=1$.  We use
  doubling trick to deal with just holomorphic functions (for further details, see Appendix A of \cite{Hatefi:2012wj}). The amplitude of two fermions and one gauge field has been found out in \cite{Polchinski:1998aa}; however, to get familiar with the notations let us start working with tree level amplitude of two fermions and one scalar field. This three point function   $<  V_{\bar\psi}V_{\psi}V_{\phi} >$ at disk level is given by
 \begin{eqnarray}
{\cal A}^{\bPsi,\Psi,\phi} & \sim & \sum_{\rm non-cyclic} \int dx_1dx_2dx_3
 \Tr\lan
V_{\bPsi}^{(-1/2)}(x_1)V_{\Psi}^{(-1/2)}(x_2)V_{\phi}^{(-1)}(x_3)\ran,\labell{cor10}\eeqa

Substituting the vertex operators and taking into account the following holomorphic correlators
\begin{eqnarray}
\lan X^{\mu}(z)X^{\nu}(w)\ran & = & -\frac{\alpha'}{2}\eta^{\mu\nu}\log(z-w) , \non \\
\lan \psi^{\mu}(z)\psi^{\nu}(w) \ran & = & -\frac{\alpha'}{2}\eta^{\mu\nu}(z-w)^{-1} \ ,\non \\
\lan\phi(z)\phi(w)\ran & = & -\log(z-w) \ .
\labell{prop}\end{eqnarray}
\reef{cor10} can be written down as
\beqa {\cal A}^{\bar\psi, \psi, \phi}& = & iT_p 2^{1/2}\pi\alpha' \int
 dx_{1}dx_{2}dx_{3}
 \xi_{1i} x_{12}^{-1/4}(x_{13}x_{23})^{-1/2}|x_{12}|^{\alpha'^2 k_1.k_2}|x_{13}|^{\alpha'^2 k_1.k_3}
 |x_{23}|^{\alpha'^2 k_2.k_3}\nonumber\\&&
\times<:S_{A}(x_1):S_{B}(x_2):\psi^i(x_3):>\bu_1^A u_2^B,\label{nop}\eeqa

Note that we have normalised the amplitude by a coefficient of $(iT_p 2^{1/2}\pi\alpha')$. Using
\beqa
 <:S_{A}(x_1):S_{B}(x_2):\psi^i(x_3):>=2^{-1/2}x_{12}^{-3/4}(x_{31}x_{32})^{-1/2}
(\gamma^{i})_{AB},\nonumber
\eeqa
 one can show that \reef{nop} is now $SL(2,R)$ invariant. We do gauge fixing as
$(x_1,x_2,x_3)=(0,1,\infty)$. Setting this gauge fixing into \reef{nop}, the amplitude becomes
 \beqa
{\cal A}^{\bPsi,\Psi,\phi} & = & iT_p \pi\alpha' \bu_1^A\ga^{i}_{AB}u_2^B\xi_{i}\left(\Tr(\lam_1\lam_2\lam_3)-\Tr(\lam_1\lam_3\lam_2)\right),\labell{amp2}
\eeqa
 The final result of the string theory can be reproduced in field theory by extracting the kinetic term of the fermion fields
 $(2\pi\alpha' T_p)\Tr(\bar\psi\ga^{\mu} D_{\mu}\psi)$. One has to extract the covariant derivative of fermion field and take into account the commutator in its definition as $(D^i\psi=\partial^i\psi-i[\phi^i,\psi])$.
 It is also good to know that the Wick-like rule \cite{Liu:2001qa,Kostelecky:1986xg} has been extended in \cite{Garousi:2007fk,Hatefi:2008ab,Hatefi:2010ik,Hatefi:2012wj} to find  various correlators  including two spin operators and an infinite number of fermions and/or currents.

\subsection{ The complete and closed form of $<V_{C} V_{\phi} V_{\bar\psi}V_{\psi} >$  }

After providing the necessary details we now compute the complete form (entire result to all  orders in $\alpha'$) of  the scattering amplitude
of one closed string RR (in the bulk), two fermions with the same chirality and a scalar field in the world volume of BPS branes.
Regarding the chirality of the fermions our computation makes sense just in IIB superstring theory and the entire result can not be extended to IIA (because fermions have different chirality in IIA).  As a matter of fact neither there are any gauge nor scalar $(t+s+u)-$ channel poles for this particular amplitude in IIA. Therefore all order corrections of two fermions-two scalars and two fermions-a gauge-a scalar of this paper can not be  applied to IIA. This $<V_{C} V_{\phi} V_{\bar\psi}V_{\psi} >$ amplitude can be looked for as follows
\begin{eqnarray}
{\cal A}^{C \phi  \bar\psi\psi} & \sim & \int dx_{1}dx_{2}dx_{3}dzd\bar{z}\,
  \lan V_{\phi}^{(0)}{(x_{1})}
V_{\bar\psi}^{(-1/2)}{(x_{2})}V_{\psi}^{(-1/2)}{(x_{3})}
V_{RR}^{(-\frac{1}{2},-\frac{1}{2})}(z,\bar{z})\ran,\labell{sstring}\eeqa

For disk amplitudes all three open strings have to be embedded on the boundary and RR should be located  inside of the disk. Notice that we just want to keep track of $\Tr(\lambda_1\lambda_2\lambda_3)$  ordering. By replacing all the vertex operators into \reef{sstring}, one might reveal that
the amplitude should be divided to two different parts $({\cal A}_1,{\cal A}_2)$. The final result is complicated so we decided to carry out each part of the amplitude separately.
\vskip.01in

First we look for $({\cal A}_1)$ in which for this part we need to know the correlation function of four spin operators (with the same chirality) in ten dimensions. This correlation function can be found in \cite{Friedan:1985ge,Hartl:2010ks},
thus we just replace it into the amplitude and the result should be read as
\beqa {\cal A}_{1}^{C \phi \bar\psi \psi}& = & \frac{\mu_p \pi^{-1/2}}{4}
 \int
 dx_{1}dx_{2}dx_{3}dx_{4} dx_{5}\,
(P_{-}\fsH_{(n)}M_p)^{\ga\delta}\xi_{1i} \bu_1^A u_2^B  (x_{23}x_{24}x_{25}x_{34}x_{35}x_{45})^{-1}
\nonumber\\&&\times \frac{1}{2}\bigg[(\ga^\mu C)_{\ga\delta}(\ga_\mu C)_{AB}x_{43} x_{52}-(\ga^\mu C)_{\ga B}(\ga_\mu C)_{A \delta}x_{45} x_{23}\bigg] I_1
 \Tr(\lam_1\lam_2\lam_3),\labell{125}\eeqa
 where we normalised the amplitude by $\frac{\mu_p \pi^{-1/2}}{4}$,
$x_{ij}=x_i-x_j, x_4=z=x+iy, x_5=\bar z=x-iy$, and
\beqa
I_1&=&{<:\partial X^i(x_1)e^{\alpha' ik_1.X(x_1)}:e^{\alpha' ik_2.X(x_2)}
:e^{\alpha' ik_3.X(x_3)}:e^{i\frac{\alpha'}{2}p.X(x_4)}:e^{i\frac{\alpha'}{2}p.D.X(x_5)}:>},
\nonumber
\eeqa
Using Wick theorem one obtains
\beqa
I_1= \bigg(\frac{ip^i  x_{54}}{x_{14} x_{15}}\bigg)  |x_{12}|^{\alpha'^2 k_1.k_2}|x_{13}|^{\alpha'^2 k_1.k_3}|x_{14}x_{15}|^{\frac{\alpha'^2}{2} k_1.p}|x_{23}|^{\alpha'^2 k_2.k_3}|
x_{24}x_{25}|^{\frac{\alpha'^2}{2} k_2.p}
|x_{34}x_{35}|^{\frac{\alpha'^2}{2} k_3.p}|x_{45}|^{\frac{\alpha'^2}{4}p.D.p},
\nonumber\eeqa
By replacing $I_1$ in the amplitude, we realise that the amplitude is SL(2,R) invariant. Using the following Mandelstam variables
\beqa
s&=&-\frac{\alpha'}{2}(k_1+k_3)^2, \quad t=-\frac{\alpha'}{2}(k_1+k_2)^2, \quad u=-\frac{\alpha'}{2}(k_3+k_2)^2,
\nonumber
\eeqa
and in particular carrying out a special gauge fixing as $(x_1=0,x_2=1,x_3=\infty)$, one can obtain the first part of the amplitude as
\beqa {\cal A}_{1}^{C \phi \bar\psi \psi}& = & \frac{\mu_p \pi^{-1/2}}{4}(P_{-}\fsH_{(n)}M_p)^{\ga\delta}\xi_{1i} \bu_1^A u_2^B (\frac{-ip^i}{2}) \int\int
 dz d\bar z |z|^{2t+2s-2}|1-z|^{2t+2u-2} (z-\bar z)^{-2(t+s+u)},
\nonumber\\&&\times  \bigg[(\ga^\mu C)_{\ga\delta}(\ga_\mu C)_{AB}(1-\bar z)+(z-\bar z)(\ga^\mu C)_{\ga B}(\ga_\mu C)_{A \delta}\bigg] \Tr(\lam_1\lam_2\lam_3)
\labell{amp3q},\eeqa
 In order to actually get the entire result, these integrals must be done on the closed string position (we propose \cite{Fotopoulos:2001pt} and the  Appendix B of \cite{Hatefi:2012wj} for further details). Therefore the complete form of the first part of the amplitude is given as follows
\beqa
{\cal A}_{1}^{C \phi \bar\psi \psi}& = & \frac{\mu_p \pi^{-1/2}}{4} (P_{-}\fsH_{(n)}M_p)^{\ga\delta}\xi_{1i} \bu_1^A u_2^B (\frac{-ip^i}{2})
\nonumber\\&&\times  \bigg[(\ga^\mu C)_{\ga\delta}(\ga_\mu C)_{AB} (st L_1+\frac{1}{2} L_2)+(\ga^\mu C)_{\ga B}(\ga_\mu C)_{A \delta} L_2\bigg] \Tr(\lam_1\lam_2\lam_3)
\labell{amp3q},\eeqa

with
\beqa
L_1&=&(2)^{-2(t+s+u)}\pi{\frac{\Gamma(-u)
\Gamma(-s)\Gamma(-t)\Gamma(-t-s-u+\frac{1}{2})}
{\Gamma(-u-t+1)\Gamma(-t-s+1)\Gamma(-s-u+1)}},\nonumber\\
L_2&=&(2)^{-2(t+s+u)+1}\pi{\frac{\Gamma(-u+\frac{1}{2})
\Gamma(-s+\frac{1}{2})\Gamma(-t+\frac{1}{2})\Gamma(-t-s-u+1)}
{\Gamma(-u-t+1)\Gamma(-t-s+1)\Gamma(-s-u+1)}}.
\label{Ls}
\eeqa

\vskip.1in

 This part of the amplitude is not vanished for different cases. For example for $n=p+2$ the first term in \reef{amp3q} has infinite singularities in u-channel and in particular it involves many contact terms.  The expansion is low energy expansion which reflects the fact that all Mandelstam variables must send to zero (for further details on the expansions see \cite{Hatefi:2010ik}).
 Hence, it becomes obvious that the first term of \reef{amp3q} includes all massless
 poles; however, we postpone its field theory computations to the next section to see what kinds of open strings, namely  gauge/ scalar or fermion can be replaced in the propagator.

 \vskip.1in
 Let us move to the second part of the amplitude.
Having replaced the second part of scalar vertex operator and the other vertices into \reef{sstring}, the second part of the amplitude can be found as follows
 \beqa {\cal A}_{2}^{C \phi \bar\psi \psi}& = & \frac{\mu_p \pi^{-1/2}}{4}\int
 dx_{1}dx_{2}dx_{3}dx_{4} dx_{5}\,
(P_{-}\fsH_{(n)}M_p)^{\ga\delta}\xi_{1i} (2ik_{1a})\bu_1^{\alpha} u_2^{\beta} ( x_{23}x_{24}x_{25}x_{34}x_{35} x_{45})^{-1/4} \nonumber\\&&\times
 <:\psi^a\psi^i(x_1):S_{\alpha}(x_2):S_{\beta}(x_3):S_{\ga}(x_4):S_{\delta}(x_5):>
 I
 \Tr(\lam_1\lam_2\lam_3),\labell{1289}\eeqa

 in which
\beqa
I=|x_{12}|^{\alpha'^2 k_1.k_2}|x_{13}|^{\alpha'^2 k_1.k_3}|x_{14}x_{15}|^{\frac{\alpha'^2}{2} k_1.p}|x_{23}|^{\alpha'^2 k_2.k_3}|
x_{24}x_{25}|^{\frac{\alpha'^2}{2} k_2.p}
|x_{34}x_{35}|^{\frac{\alpha'^2}{2} k_3.p}|x_{45}|^{\frac{\alpha'^2}{4}p.D.p},
\nonumber\eeqa

\vskip 0.1in

 The only subtlety in the second part of the amplitude is how to derive the correlation function between four spin operators (with the same chirality) and one current. Here we try to summarise the procedure of deriving this correlator. First we need to take into account the following OPE
 \beqa
 :\psi^{\mu}\psi^{\nu}(x_1):S_{\alpha}(x_2):&\sim& -\frac{1}{2}(\Gamma^{\mu\nu})_{\alpha}^{\lambda} S_{\lambda}(x_2)x_{12}^{-1}\label{esi1},
 \eeqa
with the definition of antisymmetric matrix as
 \beqa\Gamma^{\mu\nu}&=&\frac{1}{2}(\ga^\mu\ga^\nu-\ga^\nu\ga^\mu),\nonumber\eeqa
 The next step is to replacing
 this OPE \reef{esi1} into the following correlator
 \beqa
 <:\psi^a\psi^i(x_1):S_{\alpha}(x_2):S_{\beta}(x_3):S_{\ga}(x_4):S_{\delta}(x_5):>,\nonumber\eeqa
and make use of the rest of the correlator which is the correlator of four spin operators (it is given in
\cite{Hartl:2010ks} and it has two different terms). Note that we have to apply the same formalism for the other OPEs and finally add them up. Concerning this method we have eight different terms; however, in order to obtain the final answer  some extraordinary works are needed. Let us point them out.
We need to extract all gamma matrices and make use of the commutator and anti commutator relations  $\bigg\{\gamma^a,\gamma^b\bigg\}=-2\eta^{ab},\quad  \bigg\{\gamma^a,\gamma^i\bigg\}=0 $. The next step is to use the world-sheet fermion correlators as below:
\beqa
\lan \psi^{\mu}(z)\psi^{\nu}(w) \ran & = & -\frac{\alpha'}{2}\eta^{\mu\nu}(z-w)^{-1} \ ,\non \\
 <:S_{A}(x_1):S_{B}(x_2):\psi^i(x_3):>&=&2^{-1/2}x_{12}^{-3/4}(x_{31}x_{32})^{-1/2}
(\gamma^{i})_{AB}.\nonumber
\eeqa

 We also need to add all the terms carrying  common coefficients of the gamma matrices.
Finally one has to construct  different combinations of the gamma matrices.
Having taken remarks that appeared in  Appendix A.1, A.3, B.3 and section 6 of \cite{Hartl:2010ks}, one can clarify how the various terms come from. The final answer for $ <:\psi^a\psi^i(x_1):S_{\alpha}(x_2):S_{\beta}(x_3):S_{\ga}(x_4):S_{\delta}(x_5):>$ has rather complicated result, therefore let us just point out several tests in favor of getting the correct result for our correlator.

\vskip.1in

The first test of our computation is to produce the leading singularities of the amplitude in which our calculation  passes this test. The other unusual  check after having replaced the final answer for the correlator into the amplitude is in fact the SL(2,R) invariance of the amplitude in which our result  satisfies that constraint. We gauge fix the amplitude as before and evaluate the integrals on closed string location. We write down the final answer for the second part of the amplitude:
\beqa {\cal A}_{2}^{C \phi \bar\psi \psi}  & = & \frac{\mu_p \pi^{-1/2}}{16} (P_{-}\fsH_{(n)}M_p)^{\ga\delta}\xi_{1i}(2ik_{1a}) \bu_1^{\alpha} u_2^{\beta}  \bigg({\cal A}_{21}+{\cal A}_{22}+{\cal A}_{23}+{\cal A}_{24}+{\cal A}_{25}+{\cal A}_{26}\bigg)\nonumber\\&&\times\Tr(\lam_1\lam_2\lam_3),\nonumber\eeqa

such that
\beqa {\cal A}_{21} &=&-(\Gamma^{ai\mu } C)_{\alpha\beta}(\ga_\mu C)_{\ga\delta}
\bigg[\frac{1}{2} u(s+t) L_1+L_3(-s-t)\bigg],
\nonumber\\
{\cal A}_{22} &=&-(\Gamma^{ai\mu } C)_{\alpha\delta}(\ga_\mu C)_{\ga\beta}
\bigg[L_1(us)-\frac{1}{2}L_2\bigg],
\nonumber\\
{\cal A}_{23} &=& (\Gamma^{ai\mu } C)_{\ga\beta}(\ga_\mu C)_{\alpha\delta}
\bigg[L_1(ut)-\frac{1}{2}L_2\bigg],
\nonumber\\
{\cal A}_{24} &=&(\Gamma^{ai\mu } C)_{\ga\delta}(\ga_\mu C)_{\alpha\beta}
\bigg[L_1(st)+\frac{1}{2}L_2\bigg],
\nonumber\\
{\cal A}_{25} &=&(P_{-}\fsH_{(n)}M_p)^{\ga\delta}\xi_{1i}(2ik_{1a}) \bu_1^{\alpha} u_2^{\beta}
\bigg[u(s+t)L_1+L_2\bigg]\bigg((\ga^{i} C)_{\ga\beta}(\ga^a C)_{\alpha\delta}-(\ga^{i} C)_{\alpha\delta}(\ga^a C)_{\ga\beta}\bigg),
\nonumber\\
{\cal A}_{26} &=&
\bigg[-\frac{1}{2}u(s-t)L_1+(-t+s)L_3
\bigg]\bigg(-(\ga^{i} C)_{\ga\delta}(\ga^a C)_{\alpha\beta}+(\ga^{i} C)_{\alpha\beta}(\ga^a C)_{\ga\delta}\bigg)
\labell{ampc},
\eeqa
where $L_1,L_2$ are appeared in \reef{Ls} and $L_3$ is
\beqa
L_3&=&(2)^{-2(t+s+u)-1}\pi{\frac{\Gamma(-u+\frac{1}{2})
\Gamma(-s+\frac{1}{2})\Gamma(-t+\frac{1}{2})\Gamma(-t-s-u)}
{\Gamma(-u-t+1)\Gamma(-t-s+1)\Gamma(-s-u+1)}}.
\label{Ls2}
\eeqa

\vskip 0.1in

The first terms of $ {\cal A}_{21}$,$ {\cal A}_{22}$,$ {\cal A}_{25}$,$ {\cal A}_{26}$ ($ {\cal A}_{23}$) have just  t-channel (s-channel) fermion poles  and  in particular all the terms  including the coefficients of $L_2$ are just related to infinite contact interactions of one RR, two fermions and one scalar field in which they do not have any contribution to the singularities. On the other hand the first term of $ {\cal A}_{24}$ has just either infinite u-channel gauge or scalar poles.
Notice that the second terms of $ {\cal A}_{21}$,$ {\cal A}_{25}$,$ {\cal A}_{26}$ involve just s-channel  fermion poles. Finally the third and fourth terms of $ {\cal A}_{21}$,$ {\cal A}_{26}$ consist of indeed an infinite number of massless $(t+s+u)$-channel poles. In field theory  we would clarify what kinds of poles we would have.
\vskip .1in

 Note that these infinite poles have to be produced either by infinite higher derivative corrections to two scalars-two fermions or by two fermions-one scalar-one gauge field corrections in IIB. The amplitude is also antisymmetric with respect to the interchange of the fermions as we
expected.
\vskip .1in

We make some comments on T-duality. The complete form of this S-matrix can not be obtained by setting T-duality to the previous results (see\cite{Hatefi:2012ve,Hatefi:2012rx}), since  our amplitude includes some terms carrying $p^i$(closed string momentum in transverse direction). These terms can not derived by applying T-duality, given the fact that winding modes are not embedded in the explicit form of RR vertex operator. Likewise it is shown in \cite{Hatefi:2012ve} that $C\phi AA$ (one RR, a scalar and two gauge fields) amplitude can not be fully derived from $CAAA$. Thus we employ direct computations to find out some special patterns for superstring amplitudes, including fermion vertex operators.

\vskip .1in
  The expansion is low energy expansion which reflects the fact that all Mandelstam variables  should send to zero
  ($t,s,u \rightarrow 0$) such that this relation $t+s+u=-p^ap_a$ holds. The closed form of the expansion of $st L_1, L_3$ (to be able to produce u- channel poles \cite{Hashimoto:1996kf} and $(t+s+u)$ channel poles \cite{Hatefi:2010ik}) can be written down as
\beqa
 stL _1&=&-\pi^{3/2}\bigg[\sum_{n=-1}^{\infty}b_n\bigg(\frac{1}{u}(t+s)^{n+1}\bigg)+\sum_{p,n,m=0}^{\infty}e_{p,n,m}u^{p}(st)^{n}(s+t)^m\bigg],
 \nonumber\\
 L _3 &=&-\frac{\pi^{5/2}}{2}\left( \sum_{n=0}^{\infty}c_n(s+t+u)^n\right.
\left.+\frac{\sum_{n,m=0}^{\infty}c_{n,m}[s^n t^m +s^m t^n]}{(t+s+u)}\right.\nonumber\\
&&\left.+\sum_{p,n,m=0}^{\infty}f_{p,n,m}(s+t+u)^p[(s+t)^{n}(st)^{m}]\right)
\labell{highcaap}.
\eeqa
Where $su L_1$ and $ut L_1$ can be derived by replacing $t \leftrightarrow  u$ and $s\leftrightarrow u$
inside $st L_1$  expansion and $ L_2=-4(t+s+u)L_3$ which has just contact terms. Some of the coefficients are
\beqa
b_{-1}=1,\,b_0=0,\,b_1=\frac{1}{6}\pi^2,\,b_2=2\z(3),c_0=0,c_1=-\frac{\pi^2}{6},e_{0,0,1}=\frac{1}{3}\pi^2, c_2=-2\z(3),
\,c_{1,1}=\frac{\pi^2}{6},\nonumber\\
e_{0,1,0}=2\z(3),e_{1,0,0}=\frac{1}{6}\pi^2,e_{1,0,2}=\frac{19}{60}\pi^4,e_{1,0,1}=6\z(3),c_{0,0}=\frac{1}{2},
f_{0,1,0}=\frac{\pi^2}{3},f_{0,0,1}=-2\z(3).\labell{577}\eeqa
It is important to mention that the general structure of $b_n$  coefficients of this paper is exactly the same structure of  $b_n$  coefficients that appeared in the amplitude of one RR and three scalars and they have quite universal behavior \cite{Hatefi:2010ik}; however, some the coefficients of
 $c_n,f_{p,n,m}$ include differences from the coefficients of non-BPS amplitudes \cite {Hatefi:2008ab}. Let us move to field theory section and produce all  $u,t,s$-channel gauge, scalar/fermion poles. In addition to them,  we study different $(u+t+s)-$ channel scalar and gauge poles in order to obtain  all infinite higher derivative corrections of two fermions-two scalars  or two fermions-one gauge-one scalar in the world volume of BPS branes of IIB superstring theory.
\section{ Infinite $u-$channel scalar poles for $p+2=n$ case }
One can expand all the terms  of the closed form of  $<V_{C} V_{\phi} V_{\bar\psi} V_{\psi}>$     which carry $st L_1$ coefficient, to actually produce both infinite u-channel scalar and gauge poles. It is shown in \cite{Hatefi:2010ik,Hatefi:2012wj,Hatefi:2012ve} that the kinetic term of gauge and scalar fields do not  receive any correction. If one considers the $st L_1$ expansion then one can collect all the infinite u-channel scalar poles of the string theory amplitude as follows
\beqa
{\cal A}_{1}^{C \phi \bar\psi \psi} & = & \frac{\mu_p ip^i\pi}{8} \sum_{n=-1}^{\infty}b_n\bigg(\frac{1}{u}(t+s)^{n+1}\bigg)(P_{-}\fsH_{(n)}M_p)^{\ga\delta}\xi_{1i} \bu_1^A u_2^B
\nonumber\\&&\times  (\ga^\mu C)_{\ga\delta}(\ga_\mu C)_{AB}
\Tr(\lam_1\lam_2\lam_3)
\labell{amp3q11},\eeqa
In the above amplitude $\mu$ can be both world volume and transverse direction. First we set it to transverse
direction $(\mu=j)$ and extract the trace as
\beqa
(P_{-}\fsH_{(n)}M_p)^{\ga\delta}(\ga^j C)_{\ga\delta}&=&\frac{32 }{2(p+1)!} (\veps^v)^{a_0\cdots a_{p}}H^{j}_{a_0\cdots a_{p}}. \eeqa

Replacing the trace in the amplitude we obtain
\beqa
{\cal A}_{1}^{C \phi \bar\psi \psi}&=& \frac{2\mu_p ip^i\pi}{(p+1)!} \sum_{n=-1}^{\infty}b_n\bigg(\frac{1}{u}(t+s)^{n+1}\bigg)\xi_{1i} \bu_1^A (\ga_j )_{AB}  u_2^B
(\veps^v)^{a_0\cdots a_{p}}H^{j}_{a_0\cdots a_{p}} \Tr(\lam_1\lam_2\lam_3).
\labell{ampbb}\eeqa
In below we will show that the kinetic term of fermion fields has to be fixed and it does not receive any correction.
The massless poles should be reproduced by the following non-Abelian kinetic terms \footnote{By replacing $2\pi\alpha' \phi^i=X^i$ the kinetic term of the scalar field gets canonically normalised; however, in this paper we  keep the standard notation  for the kinetic term  of the open strings as they appear in \reef{kin}. }:
\beqa
-T_p (2\pi\alpha')\Tr\left(\frac{(2\pi\alpha')}{2} D_a \phi^i D^a\phi_i-\frac{(2\pi\alpha')}{4}F_{ab}F^{ba}-\bPsi\ga^aD_a\Psi\right)\labell{kin}
\eeqa
To work with the field theory of an amplitude including RR and some massless scalar fields one must consider three different approaches to explore  their vertices. Basically these methods are either Wess-Zumino (WZ) terms that introduced by  Myers \cite{Myers:1999ps}, or the needed pull-back methods or the so called  Taylor-expansion (they are argued in section 5 of \cite{Hatefi:2012wj}).

One has to take into account the following field theory amplitude to produce all scalar u-channel poles
\beqa
{\cal A}&=& V_{\alpha}^{i}(C_{p+1},\phi_1,\phi)G_{\alpha\beta}^{ij}(\phi) V_{\beta}^{j}(\phi,\bPsi_1,\Psi_2).\labell{amp549}
\eeqa
Here we should employ Taylor expansion to obtain the vertex of two scalars and one RR-$(p+1)$ form field as follows:
 \beqa
i\frac{\lambda^2\mu_p}{2!(p+1)!}\int d^{p+1}\sigma
(\veps^v)^{a_0\cdots a_{p}}\,\Tr\left(\partial_i\partial_j C^{(p+1)}_{a_0\cdots a_{p}}\phi^i\phi^j\right),\,
 \label{rr}\eeqa

where $\lambda=2\pi\alpha'$. The vertex of one RR and two scalars can be constructed from \reef{rr} 
\beqa
 V_{\alpha}^{i}(C_{p+1},\phi_1,\phi)&=&
 i\frac{\lambda^2\mu_p}{2!(p+1)!}(-ip^i) H^{j}_{a_0\cdots a_{p}} \xi_{1j}
(\veps^v)^{a_0\cdots a_{p}}\,\Tr\left(\lambda_1\lambda^{\alpha}\right).\,
 \label{rr2}\eeqa
The scalar propagator is derived from scalar fields's kinetic term (the first term in \reef{kin}). To obtain the vertex of two on-shell fermions and one off-shell scalar field  we need to work with the kinetic term of fermion fields (the last term in \reef{kin}) where the commutator in the definition of covariant derivative  of fermion field has to be considered, such that 
 \beqa
 V^{\beta}_{j}(\bPsi_1,\Psi_2,\phi)&=&T_p(2\pi\alpha')\bu_1^A\ga^j_{AB}u_2^B\left(\Tr(\lam_2\lam_3\lam^\beta)-\Tr(\lam_3\lam_2\lam^\beta)\right),\nonumber\\
 G_{\alpha\beta}^{ij}(\phi) &=&\frac{-i\delta_{\alpha\beta}\delta^{ij}}{T_p(2\pi\alpha')^2
k^2}=\frac{-i\delta_{\alpha\beta}\delta^{ij}}{T_p(2\pi\alpha')^2 u},
 \label{rr3}\eeqa

$k$ is the momentum of off-shell scalar field in the propagator. Now if we replace the above vertices in  the field theory amplitude of \reef{amp549} then the first simple u-channel scalar pole of the string theory amplitude (for $n=-1$ in \reef{ampbb}) can be precisely  produced. However, the amplitude has infinite u-channel poles. In order to deal with them  a key point has to be made.
The kinetic term of fermion fields has no correction (as it is fixed in DBI action)  and scalar propagator does not receive any correction either (because it is just simple pole). Therefore the only way to produce all the other massless scalar poles is to devote infinite higher derivative corrections to the vertex of one RR-$(p+1)$ form field and two scalar fields as follows:
 \beqa
i\frac{\lambda^2\mu_p}{2!(p+1)!}\int d^{p+1}\sigma
(\veps^v)^{a_0\cdots a_{p}}\,\sum_{n=-1}^{\infty}b_n(\alpha')^{n+1}\Tr\left(\partial_i\partial_j C^{(p+1)}_{a_0\cdots a_{p}}
D^{a_0}\cdots D^{a_n}\phi^i D_{a_0}\cdots D_{a_n}\phi^j\right).\,
 \label{rr55}\eeqa
 Notice that in \reef{rr55} all the commutators in the covariant derivative of scalar fields should be ignored as we need not have any gauge field. Now by applying standard field theory techniques we can extract the vertex of one RR, one on-shell fermion and an on-shell fermion field from \reef{rr55} as below:
 \beqa
 V_{\alpha}^{i}(C_{p+1},\phi_1,\phi)=
 i\frac{\lambda^2\mu_p}{2!(p+1)!}(-ip^i) H^{j}_{a_0\cdots a_{p}} \xi_{1j}
(\veps^v)^{a_0\cdots a_{p}}\sum_{n=-1}^{\infty}b_n (\alpha'k_1.k)^{n+1}\Tr\left(\lambda_1\lambda^{\alpha}\right),\,
 \label{rr211}\eeqa
substituting \reef{rr211} and \reef{rr3} into \reef{amp549}, we obtain 
\beqa
 \frac{\alpha'\pi p^i \mu_p}{(p+1)!} \xi_{1i} \bu_1^A (\ga_j)_{AB} u_2^B \sum_{n=-1}^{\infty}b_n \frac{(t+s)^{n+1}}{u}(\veps^v)^{a_0\cdots a_{p}}H^{j}_{a_0\cdots a_{p}} \Tr(\lam_1\lam_2\lam_3).
\labell{rr66},\eeqa
which is exactly all  the infinite $u$-channel scalar poles inside the string amplitude of \reef{ampbb}, as we expected. Hence to provide all infinite $u$-channel scalar poles in field theory, we had to generalise the vertex of one RR $(p+1)$-form field, an on-shell and an off-shell scalar field. We observed  that the closed string Ramond-Ramond has induced all infinite higher derivative corrections to two scalar fields.
This phenomenon seemed to be universal as it so happens for the other BPS and non-BPS amplitudes (see \cite{Hatefi:2010ik,Hatefi:2012ve,Hatefi:2012rx,Hatefi:2012wj}).
In the next section we construct all infinite higher derivative corrections of  one RR-$(p-1)$ form field, one gauge and one scalar field to be able to produce all the infinite $u$-channel gauge poles of  $C \phi \bar\psi\psi$ amplitude.
\section{ Infinite $u-$channel gauge poles for $p=n$ case}
 In this section we are going to produce all the singularities that appeared in ${\cal A}_{24}$.  By extracting  the trace, replacing the first part of the expansion of $st L_1$(just singularities) and setting $\mu$ to be located in world volume direction ($\mu=b$)  in ${\cal A}_{24}$,
 one can express all the infinite $u$-channel gauge poles  of the string theory amplitude as follows:
\beqa
{\cal A}_{24} &=& \frac{4\mu_p \pi\xi_{1i}(ik_{1a}) }{p!} \bu_1^{A} (\ga_b)_{AB}  u_2^{B}  \sum_{n=-1}^{\infty}b_n \frac{1}{u}(t+s)^{n+1}\nonumber\\&&\times
(\veps^v)^{a_0\cdots a_{p-2}ab}H^{i}_{a_0\cdots a_{p-2}}\Tr(\lam_1\lam_2\lam_3).
 \label{mm1}\eeqa

Note that we have ignored the second term of ${\cal A}_{24}$ because it  is just contact interaction which has no singularity. In \cite{Hatefi:2010ik,Hatefi:2012wj,Hatefi:2012ve} we have explained how to derive all contact terms in the field theory. To produce all the infinite $u$-channel gauge poles the following Feynman rule in field theory has to be considered

 \beqa
{\cal A}&=& V_{\alpha}^{a}(C_{p-1},\phi_1,A)G_{\alpha\beta}^{ab}(A) V_{\beta}^{b}(A,\bPsi_1,\Psi_2),\labell{amp548}
\eeqa

where $V_{\alpha}^{a}(C_{p-1},\phi_1,A)$ can be derived by making use of the combination of Taylor expansion and WZ terms as follows
 \beqa
i\frac{\lambda^2\mu_p}{p!}\int d^{p+1}\sigma
\Tr\left(\partial_i C_{(p-1)}\wedge F \phi^i\right).\,
 \label{mm33}\eeqa
 
If we apply standard field theory techniques then one can feasibly derive the vertex of one RR-$(p-1)$-form field , one off-shell gauge field and one on-shell scalar field as below
\beqa
 V_{\alpha}^{a}(C_{p-1},\phi_1,A)&=&
 i\frac{\lambda^2\mu_p}{p!} H^{i}_{a_0\cdots a_{p-2}} \xi_{1i} k_{a_{p-1}}
(\veps^v)^{a_0\cdots a_{p-1}a}\,\Tr\left(\lambda_1\lambda^{\alpha}\right)\,
 \label{mm44}\eeqa
where k is momentum of off-shell gauge field $k=k_2+k_3$.
To derive gauge propagator one should work with the kinetic term of gauge fields. To obtain
the vertex of two on-shell fermion fields and one off-shell gauge field $(V_{\beta}^b(\bPsi_1,\Psi_2, A))$, one needs to make use of the kinetic term of fermions and also to extract covariant derivative of the fermion field $ 
(D^a\psi= \partial^a\psi-i[A^a,\psi])$. Essentially one has to take into account all possible orderings of the gauge and fermions to be able to  obtain  $V_{\beta}^{b}(A,\bPsi_1,\Psi_2)$ as well as gauge propagator
 \beqa
 V_{\beta}^b(A,\bPsi_1,\Psi_2 )&=&T_p(2\pi\alpha')\bu_1^A\ga^b_{AB}u_2^B\left(\Tr(\lam_2\lam_3\lam^\beta)-\Tr(\lam_3\lam_2\lam^\beta)\right),\nonumber\\
 G_{\alpha\beta}^{ab}(A) &=&\frac{-i\delta_{\alpha\beta}\delta^{ab}}{T_p(2\pi\alpha')^2
k^2}=\frac{-i\delta_{\alpha\beta}\delta^{ab}}{T_p(2\pi\alpha')^2 u},
 \label{mm55}\eeqa

with $k$  becomes off-shell gauge field propagator ($k^a=(k_2+k_3)^a=(-p-k_1)^a$).
Replacing the above vertices in  the field theory amplitude of \reef{amp548} we can precisely find the first simple $u$-channel gauge pole of the string theory (for $n=-1$ in the amplitude of \reef{mm1}). The constraint for Ramond-Ramond
$p^a(\veps^v)^{a_0\cdots a_{p-1}a}=0$
has also been used. It is worth trying to produce all infinite $u$-channel massless gauge poles. We made some points in the previous section and understood that the kinetic term of fermion fields has no correction and simple scalar/gauge propagators also do not receive any correction. Hence, the only way to produce all the other $u$- channel gauge poles is to propose higher derivative corrections to the vertex of one RR-$(p-1)$-form field, one on-shell scalar and an off-shell gauge field  as follows
 \beqa
i\frac{\lambda^2\mu_p}{p!}\int d^{p+1}\sigma
 \,\sum_{n=-1}^{\infty}b_n(\alpha')^{n+1}\Tr\left(\partial_i C_{(p-1)} \wedge
D^{a_0}\cdots D^{a_n}F D_{a_0}\cdots D_{a_n}\phi^i\right).\,
 \label{mm55}\eeqa
 Now we are ready to extract the higher extension of the vertex of  $V_{\alpha}^{a}(C_{p-1},\phi_1,A)$ as below
 \beqa
 V_{\alpha}^{a}(C_{p-1},\phi_1,A)&=&
 i\frac{\lambda^2\mu_p}{p!} H^{i}_{a_0\cdots a_{p-2}} \xi_{1i} k_{a_{p-1}}
(\veps^v)^{a_0\cdots a_{p-1}a} \sum_{n=-1}^{\infty}b_n (\alpha'k_1.k)^{n+1} \Tr\left(\lambda_1\lambda^{\alpha}\right).\,
 \label{mm66}\eeqa

Notice that \reef{mm66} is all order extension of \reef{mm44}.  Substituting \reef{mm66} into \reef{amp548} we are able to exactly reproduce all infinite $u$-channel gauge poles of the amplitude that
appeared in \reef{mm1}.
Hence  closed string Ramond-Ramond $(p-1)$-form field has proposed all infinite higher derivative corrections to an on-shell scalar and one off-shell gauge field. We might wonder about this universal behavior of
higher derivative corrections that RR has induced to open strings.  Just for the completeness we mention that, this phenomenon has also been seen in non-BPS brane systems (see \cite{Hatefi:2012wj}).
\section{Infinite $t,s$-channel fermion poles }
For $C \phi \bar\psi\psi$ amplitude, there is no graviton propagator in $s,t$ channels, because the particle exchanged must have non-zero fermion number.  Furthermore, the particles exchanged are corresponding to open string excitations thus for these channels there is no coupling between one Ramond vertex operator, one scalar and one graviton as one can not saturate the total super ghost charge for disk amplitude. Therefore the only propagator for these channels is indeed fermionic propagator. To find all the singularities related to $t$-channel we need to replace just the first term of $us L_1$ expansion inside
$A_{21},A_{22},A_{25},A_{26}$, extract the related traces and simplify the amplitude more. By applying these points, we are able to write down
all $t$,($s$)-channel fermion poles of the string amplitude as below:
\beqa
{\cal A} &=& \frac{\alpha'\mu_p \pi\xi_{1i}(2ik_{1a}) }{(p+1)!} \bu_1^{A} (\ga^a)_{AB}  u_2^{B}  \sum_{n=-1}^{\infty}b_n \bigg[ \frac{1}{t}(u+s)^{n+1}-\frac{1}{s}(u+t)^{n+1}\bigg]\nonumber\\&&\times
(\veps^v)^{a_0\cdots a_{p}}H^{i}_{a_0\cdots a_{p}}\Tr(\lam_1\lam_2\lam_3).
 \label{nmm1}\eeqa
The amplitude is totally anti symmetric under interchange of the fermionic strings $(s\leftrightarrow t)$, thus there is no need to produce all s-channel poles. We just produce all fermionic t-channel poles and finally by interchanging $2\leftrightarrow 3$ and $s\leftrightarrow t$ in all kinematic relations, all s-channel fermionic poles can be concluded as well. All the terms that include $L_2$ coefficients are just contact interactions and have nothing to do with these singularities. Let us write down the rule to derive all the fermionic t-channel poles
\beqa
{\cal A}&=& V_{\alpha}(C_{p+1}, \Psi_3,\bar\Psi)G_{\alpha\beta}(\Psi) V_{\beta}(\Psi,\bar\Psi_2,\phi_1),\labell{amp560}
\eeqa
The fermion propagator is found by making use of the kinetic term of fermion fields (the last term in \reef{kin}). If we extract the covariant derivative of fermion inside its kinetic term ( $D^i\psi= \partial^i\psi-i[\phi^i,\psi]$) and count all possible orderings of the scalar and fermions, we can derive the vertex of one on-shell, one off-shell fermion and an on-shell scalar field
 ($V_{\beta}(\Psi,\bar\Psi_2,\phi_1)$) as well as fermion propagator as below
 \beqa
 V_{\beta}(\Psi,\bar\Psi_2,\phi_1)&=&T_p(2\pi\alpha')\bu_1^A\ga^j_{A}\xi_{1j}\left(\Tr(\lam_1\lam_2\lam^\beta)-\Tr(\lam_2\lam_1\lam^\beta)\right),\nonumber\\
 G_{\alpha\beta}(\psi) &=&\frac{-i\delta_{\alpha\beta}}{T_p(2\pi\alpha')
\fsk}=\frac{-i\delta_{\alpha\beta}\ga^a(k_1+k_2)_a}{T_p(2\pi\alpha') t}.
 \label{mm55}\eeqa
In order to find  $V_{\alpha}(C_{p+1},\bar \Psi,\Psi)$, one has to find out  new coupling between one RR $(p+1)-$ form field, one on-shell and one off-shell fermion field  in the world volume of BPS branes as follows
 \beqa
i\frac{(2\pi\alpha')\mu_p}{(p+1)!}\int d^{p+1}\sigma
\Tr\left( C_{a_0\cdots a_{p}} \bPsi \ga^j \partial_j \Psi \right) (\veps^v)^{a_0\cdots a_{p}}.\,
\label{nn22}\eeqa
which is in fact supersymmtrized version of  known bosonic couplings. Note that the equations of motion for fermions must be taken into account $(\fsk_{2a}\bu=\fsk_{3a}u=0 )$ as well.
Now the vertex of one RR $(p+1)$-form field, one off-shell and one on-shell fermion field can be derived from \reef{nn22} as follows 
\beqa
 V_{\alpha}(C_{p+1},\bar \Psi,\Psi_2)&=&i\frac{(2\pi\alpha')\mu_p}{(p+1)!}
 H^i_{a_0\cdots a_{p}} \ga^i u_2
(\veps^v)^{a_0\cdots a_{p}}\,\Tr\left(\lambda_3\lambda^{\alpha}\right).\,
 \label{nn33}\eeqa
If we replace the vertices that appeared in \reef{mm55} and \reef{nn33}  to the field theory amplitude of \reef{amp560} then we can find the first simple t-channel fermion pole of the string theory amplitude  (for $n=-1$ in \reef{nmm1}).
Once more in order to derive all infinite t- or s-channel fermion poles, one has to impose the infinite higher derivative interactions to new coupling \reef{nn22}. Thus the kinetic term of fermion field should be fixed. Now one has to look for  the infinite higher derivative corrections to the vertex of  $V_{\alpha}(C_{p+1},\Psi_3,\bar \Psi)$ as follows
\beqa
i\frac{(2\pi\alpha')\mu_p}{(p+1)!}\int d^{p+1}\sigma
\sum_{n=-1}^{\infty}b_n(\alpha')^{n+1}\Tr\left( C_{a_0\cdots a_{p}} D^{a_0}\cdots D^{a_n} \bPsi \ga^i D_{a_0}\cdots D_{a_n} \partial_i \Psi \right)\,(\veps^v)^{a_0\cdots a_{p}}.
\label{nn55}\eeqa
 
 If we consider \reef{nn55} then the complete form of the $V_{\alpha}(C_{p+1},\Psi_3,\bar \Psi)$ to all orders in $\alpha'$  will become 
 \beqa
 V_{\alpha}(C_{p+1},\bar \Psi,\Psi)&=&i\frac{(2\pi\alpha')\mu_p}{(p+1)!} H^i_{a_0\cdots a_{p}} \ga^i u_2
(\veps^v)^{a_0\cdots a_{p}}\,\Tr\left(\lambda_3\lambda^{\alpha}\right)\,\sum_{n=-1}^{\infty}b_n (\alpha'k_3.k)^{n+1}.
 \label{nn66}\eeqa
An important point should be emphasised, namely all the commutator terms inside the covariant derivative of fermion fields in \reef{nn55} must be neglected. Note that in producing all fermion poles in field theory side, we have employed the equations of motion for fermion fields $\fsk_2 \bu=\fsk_3 u=0$ as well. Replacing  \reef{nn66} into \reef{amp560}, one can show that  all  fermion t(s)-channel fermion poles are exactly reconstructed. Therefore the closed string Ramond-Ramond $(p+1)$-form field has induced an infinite number of  higher derivative corrections to two fermions,  two scalar fields as well as one scalar- one gauge field. This clearly confirms that, this phenomenon (producing all poles of the string amplitude by postulating infinite higher derivative corrections to the vertex of one RR and some open string vertex operators)  is quite universal and might be useful for deriving all the singularities of the higher point functions of either BPS or non-BPS branes without the need for knowing the exact results of the world sheet integrals of the higher point functions.
\subsection{Infinite higher derivative corrections to two fermions- two scalars and all scalar poles for $p=n-2$ case }
 The first goal of this section is to produce poles order by order and to actually observe whether or not the universal  conjecture on all order  $\alpha'$ higher derivative
 corrections (which has been made for bosonic amplitude in \cite{Hatefi:2012rx}) can be held for the fermionic amplitudes.  Another goal would be determining all the infinite higher derivative corrections of two fermions-two scalars in IIB superstring theory. We first consider  the infinite massless scalar  
 $(s+t+u)=-p^a p_a$  channel poles of the  string amplitude of $C_{p+1}\bar\Psi\Psi\phi$ . If we extract the related traces in the second term of $A_{26}$ then one finds out all the singularities as below
\beqa {\cal A} &=&- \frac{ \alpha'\pi^{-1/2}\mu_p}{(p+1)!} (\veps^v)^{a_0\cdots a_{p}} H^i_{a_0\cdots a_{p}}\xi_{1i}(2ik_{1a}) \bu_1^{A} (\ga^a )_{AB} u_2^{B}
\Tr\left(\lambda_1\lambda_2\lambda_3\right)\,
\bigg[
(-t+s)L_3\bigg]
 \labell{ampcmm},
\eeqa
 The  expansion of $L_3$ has infinite $(t+s+u)$- channel poles so let us see its expansion before the simplification:
 \beqa
 L _3 &=&-\pi^{5/2} \bigg(\frac{1}{2(t+s+u)}+\frac{\pi^2(t^2+s^2+u^2)}{12(t+s+u)}+\frac{\xi(3)(t^3+s^3+u^3+tsu)}{(t+s+u)}+\cdots\bigg)
\labell{highcaapbb}.
\eeqa
 For the moment we just work with the first $(t+s+u)$- channel pole. Reminding the coupling of one RR and one scalar field  is worthwhile for the field theory amplitude
\beqa
\lambda\mu_p\int d^{p+1}\sigma {1\over (p+1)!}
(\veps^v)^{a_0\cdots a_{p}}\,\Tr\left(\phi^i\right)\,
H^{(p+2)}_{ia_0\cdots a_{p}}, \label{cvb}\eeqa
In order to produce the first $(t+s+u)-$ channel scalar pole, the following Feynman  rule  has to be taken into account
\beqa
{\cal A}&=&V_{\alpha}^{i}(C_{p+1},\phi)G_{\alpha\beta}^{ij}(\phi)V_{\beta}^{j}(\phi,\bar\Psi,\Psi,\phi_1),\label{amp54m9}
\eeqa
where the scalar propagator can be readily derived from the kinetic term of the scalar fields and $V_{\alpha}^{i}(C_{p+1},\phi)$ is derived from  \reef{cvb} so that
\beqa
G_{\alpha\beta}^{ij}(\phi) &=&\frac{-i\delta_{\alpha\beta}\delta^{ij}}{T_p(2\pi\alpha')^2
k^2}=\frac{-i\delta_{\alpha\beta}\delta^{ij}}{T_p(2\pi\alpha')^2
(t+s+u)},\nonumber\\
V_{\alpha}^{i}(C_{p+1},\phi)&=&i(2\pi\alpha')\mu_p\frac{1}{(p+1)!}(\veps^v)^{a_0\cdots a_{p}}
 H^{i}_{a_0\cdots a_{p}}\Tr(\lambda_{\alpha}).
\labell{Fey}
\eeqa
Notice that scalar in  $V_{\alpha}^{i}(C_{p+1},\phi)$ must be Abelian so $\Tr(\lambda_{\alpha})$ makes sense for the Abelian matrix $\lambda_{\alpha}$. If one looks at \reef{ampcmm} then one understands that the first simple $(t+s+u)$- channel scalar pole has to be discovered by all the couplings between two scalars and two fermions in such a way that they should carry three momenta. The reason for this sharp result is that in  \reef{ampcmm} apart from
the field strength of RR (which is absorbed in $V_{\alpha}^{i}(C_{p+1},\phi)$), the other terms carry three momenta.
 One can show that in field theory analysis  the kinematic factor of $(-t+s)$  can be factorized. This is the key point in favor of a given universal conjecture of all infinite $\alpha'$ higher derivative corrections of the string amplitudes. Further details can be found in  \cite{Hatefi:2012rx}. In order to find out two fermion two scalar couplings  we need to write down  all possible desired couplings carrying three momenta as below 
\beqa
\frac{T_p (2\pi\alpha')^3}{4}(\bPsi \ga^a D_b\Psi D^a\phi^i D^b\phi_i+D^a\phi^i D^b\phi_i\bPsi \ga^a D_b\Psi  ),\quad\quad\frac{T_p (2\pi\alpha')^3}{8}\bPsi \ga^a D_a\Psi D^b\phi^i D_b\phi_i.
\label{mkcx} \eeqa
Based on the prescription  for all order higher order corrections given in \cite{Hatefi:2012rx}, one needs to consider the multiplication of the kinetic terms of the open strings to end up with their all order $\alpha'$ higher derivative corrections. Thus for this part of the amplitude, we must multiply the kinetic terms of the fermion fields and scalar fields which considered in \reef{mkcx}.
Note that the coefficients of the couplings in \reef{mkcx} would be fixed in such a way that the first $(t+s+u)$- channel scalar pole in \reef{ampcmm} would be resulted. Likewise the last section, all the  commutator terms in the definitions of the covariant derivative of fermion/scalar fields must be overlooked.
In order to produce the field theory vertex operators for the following coupling
\beqa
\bPsi \ga^a D_b\Psi D^a\phi^i D^b\phi_i,\nonumber\eeqa
one has to consider two possible orderings
 \beqa\Tr(\lam_2\lam_3\lam_1\lam_\beta), \quad\quad\Tr(\lam_2\lam_3\lam_\beta\lam_1),\nonumber\eeqa
 where $\lam_\beta$ is related to Abelian scalar field in the propagator.  By Replacing these orderings to the Feynman amplitude \reef{amp54m9}, $\Tr(\lam_1\lam_2\lam_3)$ is produced.
 Thus by extracting $\frac{T_p (2\pi\alpha')^3}{4}(\bPsi \ga^a D_b\Psi D^a\phi^i D^b\phi_i+D^a\phi^i D^b\phi_i\bPsi \ga^a D_b\Psi  )$ couplings of \reef{mkcx} and by considering all their orderings one obtains
\beqa
V_{\beta}^{j}(\phi,\bar\Psi,\Psi,\phi_1)&=&i\frac{T_p (2\pi\alpha')^3}{4} \bar u^A (\ga^a)_{AB} u^B \xi_{1j}\bigg( -k_{1a}\frac{t}{2}-k_{4a}\frac{s}{2}\bigg)\Tr(\lam_1\lam_2\lam_3\lam_\beta)
\nonumber\eeqa
Note that in order to  derive  $V_{\beta}^{j}(\phi,\bar\Psi,\Psi,\phi_1)$, we used the momentum conservation $(k_1+k_2+k_3+k_4)^a=0$ and made use of the equations of motion for fermion fields $(k^{a}_2 \gamma_a \bar u=k^{a}_3 \gamma_a  u=0)$. Setting these remarks, we are able to find the vertex of two on-shell fermions, one on-shell and one off-shell scalar field as follows
\beqa
V_{\beta}^{j}(\phi,\bar\Psi,\Psi,\phi_1)&=& i\frac{T_p (2\pi\alpha')^3}{4} k_{1a} \bar u^A (\ga^a)_{AB} u^B \xi_{1j} (-\frac{t}{2}+\frac{s}{2}) \Tr(\lam_1\lam_2\lam_3\lam_\beta).
\label{eese}\eeqa
Now if we replace \reef{eese} into  \reef{amp54m9} and also consider the first term of \reef{highcaapbb} to \reef{ampcmm} then we are able to exactly produce the first $ (t+s+u)$-channel scalar pole of \reef{ampcmm}.
\vskip.1in

For $(\bPsi \ga^a D_a\Psi D^b\phi^i D_b\phi_i )$ coupling one has to  consider two different mentioned  orderings. If one extracts this coupling and considers standard field theory techniques then  we may obtain 
 \beqa
V_{\beta}^{j}(\phi,\bar\Psi,\Psi,\phi_1)&=& -i k_{3a} \bar u^A (\ga^a)_{AB} u^B \xi_{1j} u \Tr(\lam_1\lam_2\lam_3\lam_\beta),
\nonumber\eeqa
one might apply on-shell condition $(t+s+u=0)$ to the above vertex operator to remove  Mandelstam variable of $u$;however,  by applying the equation of motion for fermion field $(\fsk_3 u^B=0)$ to the above vertex, we realise that, this coupling does not have any contribution to field theory amplitude. It is clear from the expansion of $L_3$ \reef{highcaapbb} that the string amplitude \reef{ampcmm} has infinite massless $(t+s+u)$-channel scalar poles \cite{Hashimoto:1996kf}.
It is also seen in \cite{Hatefi:2012wj} that the vertex of
$V_{\alpha}^{i}(C_{p+1},\phi)$ and the simple scalar propagator do not require any corrections, so one expects that all infinite $(t+s+u)$-channel
 scalar poles are related to
all order $\alpha'$ higher derivative corrections of  two fermions-two scalar fields of IIB superstring theory.
 \vskip.1in

Let us now generalise our method to find out all order  $\alpha'$  higher derivative corrections to two fermion-two scalar field couplings to be able to produce all the infinite $(t+s+u)$-channel scalar poles. Indeed we need to apply some higher derivative operators $(\cD_{nm},\cD'_{nm})$ to all couplings that  have non-zero contributions to the field theory amplitude (the first and second term in \reef{mkcx}). Having taken the following couplings \footnote{Recent computations for all order $\alpha'$ higher derivative corrections of two fermions-two tachyons confirm that the same universal conjecture holds even for non-BPS branes (see \cite{Hatefi:2013mwa}).}
\beqa
&&{\cal L}^{n,m}= \pi^3\alpha'^{n+m+3}T_p\bigg(a_{n,m}\Tr\bigg[\cD_{nm}\left(\bPsi \ga^a D_b\Psi D^a\phi^i D^b\phi_i \right)+\cD_{nm}\left( D^a\phi^i D^b\phi_i  \bPsi \ga^a D_b\Psi \right)
\nonumber\\&&+h.c \bigg]
+i b_{n,m}\Tr\bigg[\cD'_{nm}\left(\bPsi \ga^a D_b\Psi D^a\phi^i D^b\phi_i \right)+\cD'_{nm}\left(
 D^a\phi^i D^b\phi_i  \bPsi \ga^a D_b\Psi \right)+h.c.\bigg]
\bigg),\labell{Lnm}
\eeqa
 with the following definitions of the higher derivative operator  of $\cD_{nm} ,\cD'_{nm}$ 
\beqa
\cD_{nm}(EFGH)&\equiv&D_{b_1}\cdots D_{b_m}D_{a_1}\cdots D_{a_n}E  F D^{a_1}\cdots D^{a_n}GD^{b_1}\cdots D^{b_m}H,\nonumber\\
\cD'_{nm}(EFGH)&\equiv&D_{b_1}\cdots D_{b_m}D_{a_1}\cdots D_{a_n}E   D^{a_1}\cdots D^{a_n}F G D^{b_1}\cdots D^{b_m}H,\nonumber
\eeqa
we are now able to show that all the infinite $(t+s+u)$-channel scalar  poles can be produced.
In order to do so, we focus on the terms that carry $a_{n,m}$ coefficients.
 If we apply standard field theory techniques, consider hermitian conjugate of the first and second  couplings in \reef{Lnm}, take into account momentum conservation in world volume direction, use the equations of motion for fermion fields and finally take all different possible  orderings\footnote{Notice that BPS branes do not carry Chan -Paton factor so we do not expect to have any $(-\cD_{nm}),(-\cD'_{nm})$ operators in the couplings \reef{Lnm}.}
 then one obtains the vertex of two on-shell fermions- one on-shell and one off-shell scalar to all orders of $\alpha'$ as follows:
\beqa
V_{\beta}^{j}(\phi,\bar\Psi,\Psi,\phi_1)= i T_p\frac{ (2\pi\alpha')^3}{4} k_{1a} \bar u^A (\ga^a)_{AB} u^B \xi_{1j} \bigg(-\frac{t}{2} t^n s^m+ \frac{s}{2} t^ms^n\bigg) \Tr(\lam_1\lam_2\lam_3\lam_\beta).
\label{eesenn23}\eeqa

Now if one replaces \reef{eesenn23} into  \reef{amp54m9} then one is able to produce  exactly all the infinite $(t+s+u)$-channel scalar poles in \reef{ampcmm}. For instance by putting $n,m=0$ inside  \reef{eesenn23}, we showed (compare \reef{eesenn23} with \reef{eese}) that the first $(t+s+u)$-channel  scalar pole is produced. Let us go on, now  by replacing $n=1,m=0$ to \reef{eesenn23}  we obtain
\beqa
V_{\beta}^{j}(\phi,\bar\Psi,\Psi,\phi_1)= i T_p\frac{ (2\pi\alpha')^3}{4} k_{1a} \bar u^A (\ga^a)_{AB} u^B \xi_{1j} \bigg(-\frac{t^2}{2} + \frac{s^2}{2}\bigg),
\label{eesenn}\eeqa

let us  replace \reef{eesenn} into the field theory amplitude \reef{amp54m9} and also consider the second term of the $L_3 $ expansion of (17) inside \reef{ampcmm}. To do so, we discover that the over all coefficient of string amplitude ( $k_{1a}(s-t)$ ) can be extracted from \reef{eesenn}. The rest of the coefficients in field theory amplitude, namely  $\frac{1}{2} (s+t)$ would be cancelled if we would compare them with the coefficient of $c_{1,1}(s+t) $ of the string amplitude. Therefore the first simple $(t+s+u)$ channel scalar pole of the string and field theory amplitude is exactly matched.
If we concentrate on the terms in \reef{Lnm} carrying $b_{n,m}$ coefficients  and extract the vertex we  derive
\beqa
V_{\beta}^{j}(\phi,\bar\Psi,\Psi,\phi_1)= i T_p\frac{ (2\pi\alpha')^3}{4} k_{1a} \bar u^A (\ga^a)_{AB} u^B \xi_{1j} \bigg(-\frac{t}{2} u^n s^m+ \frac{s}{2} t^m  u^n\bigg) \Tr(\lam_1\lam_2\lam_3\lam_\beta).
\label{eesenn2}\eeqa
However, one has to apply on-shell condition $(s+t+u)=0$ into the above vertex to be able to produce all the infinite scalar poles. Indeed similar checks for all order $\alpha'$ higher derivative corrections to the other amplitudes in \cite{Hatefi:2010ik,Hatefi:2012ve} have been carried out.
This ends our goal of exploring  all order $\alpha'$ higher derivative corrections to two  fermions-two scalars of the world volume of BPS branes in  type IIB superstring theory.
\subsection{The higher derivative corrections to two fermions -one gauge and one scalar field  and the
  infinite  gauge poles for $p=n$ case }
The final singularities in the amplitude of $C_{p-1}\bar\Psi\Psi\phi$ are related to all infinite massless $(s+t+u)$-channel gauge poles of the string theory. The goal of this section is to  find out  non zero couplings of one gauge-one scalar and two fermions of IIB string theory and to fix their coefficients  by producing all infinite gauge poles of the string amplitude. Of course, essentially  one wants to derive all the infinite corrections. It would also be nice to see whether or not the universal  conjecture about higher derivative
 corrections to all orders in $\alpha'$, made in \cite{Hatefi:2012rx} holds for fermionic amplitude?
 Our last  goal is  to fix the coefficients of all order  corrections. In fact these coefficients  must be found just by comparing the couplings in field theory with string theory amplitude and not by any other tools  such as T-duality.
One  finds the singular terms in the string amplitude for $p=n$ case as follows:
\beqa {\cal A} &=&- \frac{ \alpha'\pi^{-1/2}\mu_p}{(p)!} (\veps^v)^{a_0\cdots a_{p-1}a} H_{a_0\cdots a_{p-1}}\xi_{1i}(2ik_{1a}) \bu_1^{A} (\ga^i )_{AB}u_2^{B}
\Tr\left(\lambda_1\lambda_2\lambda_3\right)\,
\bigg[-2t L_3\bigg].
 \labell{ampc33}
\eeqa
A crucial remark is in order. The vertex of RR, looks like to the fermions vertex operators
  so one may suppose  the prescription for
 $<V_{C}V_{A} V_AV_A>$  in \cite{Hatefi:2010ik}
 or $<V_{C}V_{\phi} V_AV_A>$ in \cite{Hatefi:2012ve}\footnote{The complete
form of two gauge-two scalar couplings to all orders
 in $\alpha'$ is found by comparing  field theory amplitude with
 string theory amplitude of $<V_{C}V_{\phi} V_AV_A>$ in \cite{Hatefi:2012ve}.} can be applied to this section as well but in this section we  are looking for two fermions-one gauge-one scalar couplings. Thus we are not allowed to make use of the kinetic term of the gauge fields because we need to take into account one scalar/gauge couplings to the other fields. Indeed in $<V_{C} V_{\phi} V_{\bar\psi}V_{\psi} >$  there is no external
 gauge field so these one scalar/gauge-two fermion couplings have to be searched   just by
comparison the field theory vertices with
string theory S-matrix. Eventually all the coefficients of the field theory couplings
must be fixed through comparisons field theory couplings with S-matrix elements.
\vskip.1in
 We have already pointed out that the expansion of  $L_3$  had infinite $(t+s+u)$- channel poles.
 Let us  first carry out field theory computations to produce just the first simple $(t+s+u)$- channel gauge pole.
First of all we need to have the coupling between one RR $(p-1)-$ form field and one gauge field as follows
\beqa
i(2\pi\alpha')\mu_p\int d^{p+1}\sigma {1\over (p)!}
C_{p-1}\wedge F. \label{mmnb1}\eeqa
This coupling is derived in \cite{Hatefi:2010ik}. In the  amplitude of one RR and three gauge fields \cite{Hatefi:2010ik}, we saw that $V(C_{p-1},A)$ does not receive any correction. Hence, for producing all the infinite gauge poles for $p=n$ case, one expects that the only corrections are related to the corrections  of one off-shell gauge, one on-shell scalar field and two fermion fields  (to all orders in $\alpha'$ in IIB superstring theory). Thus  the infinite higher derivative corrections  should be explored by matching all the infinite gauge poles of the string amplitude of $C_{p-1}\bar\Psi \Psi \phi$  with field theory amplitude.
The  Feynman  rule  is
\beqa
{\cal A}&=&V_{\alpha}^{a}(C_{p-1},A)G_{\alpha\beta}^{ab}(A)V_{\beta}^{b}(A,\bar\Psi,\Psi,\phi_1),\label{amp5211}
\eeqa
where the gauge propagator is obtained from the kinetic term of gauge fields and $V_{\alpha}^{a}(C_{p-1},A)$ is derived from \reef{mmnb1} such that
\beqa
G_{\alpha\beta}^{ab}(A) &=&\frac{-i\delta_{\alpha\beta}\delta^{ab}}{T_p(2\pi\alpha')^2
k^2}=\frac{-i\delta_{\alpha\beta}\delta^{ab}}{T_p(2\pi\alpha')^2
(t+s+u)},\nonumber\\
V_{\alpha}^{a}(C_{p-1},A)&=&i(2\pi\alpha')\mu_p\frac{1}{(p)!}(\veps^v)^{a_0\cdots a_{p-1}a}
 H^{}_{a_0\cdots a_{p-1}}\Tr(\lambda_{\alpha}).
\labell{Fey33}
\eeqa
The gauge field in  $V_{\alpha}^{a}(C_{p-1},A)$ must be Abelian so $\Tr(\lambda_{\alpha})$ makes sense just for the Abelian matrix $\lambda_{\alpha}$. By looking at \reef{ampc33} one understands that the first simple $(t+s+u)$-scalar pole has to be discovered with the couplings between an on-shell scalar, one off-shell  gauge and two on-shell fermions such that they have to carry three momenta, (for more explanations see the previous section). Consider the following couplings:
\beqa
\frac{T_p (2\pi\alpha')^3}{4}\bigg[\bPsi \ga^i D_b\Psi D^a\phi_i F_{ab}+\bPsi \ga^i D_b\Psi F_{ab} D^a\phi_i\bigg], \label{zzxc}
 \eeqa
In order to have general covariance in \reef{zzxc}, one has to consider the multiplications of the kinetic term  of the fermions, the field strength of gauge field and the covariant derivative of the scalar field. Notice that  the connections (commutator terms) in the definitions of the covariant derivative of fermion field  and scalar field  must be overlooked.
Let us work out \reef{zzxc}. Unlike the previous section, the only possible orderings for the first and second term of \reef{zzxc} accordingly are
$\Tr(\lam_2\lam_3\lam_1\lam_\beta),  \Tr(\lam_2\lam_3\lam_\beta\lam_1)$
where $\lam_\beta$ is related to Abelian gauge field in the propagator. To obtain the vertex of two on-shell fermions-one off-shell gauge and one on-shell scalar field, one has to extract couplings \reef{zzxc}, apply
 momentum conservation along the world volume of brane and make use of the equations of motion for fermion fields, such that 
\beqa
V_{\beta}^{b}(A,\bar\Psi,\Psi,\phi_1)&=& i\frac{T_p (2\pi\alpha')^3}{2} \bar u^A (\ga^j)_{AB} u^B \xi_{1j}\bigg( k_{1b}\frac{t}{2}\bigg)\Tr(\lam_1\lam_2\lam_3\lam_\beta).
\label{eesddem}\eeqa
Now if we replace \reef{eesddem} into \reef{amp5211} and consider the first term of the expansion of $L_3$  ( which appeared in \reef{highcaapbb})  inside \reef{ampc33} then we are able to exactly produce the first $(t+s+u)$-channel gauge pole of \reef{ampc33}.
\vskip.1in
 It is clear from  \reef{highcaapbb} that the  string amplitude \reef{ampc33} has infinite massless $(t+s+u)$-channel gauge poles. Keeping in mind  that $V_{\alpha}^{a}(C_{p-1},A)$ and the simple gauge propagator do not require any corrections, one expects that all infinite gauge poles are related to higher derivative corrections of  two fermions, one scalar and one gauge field of IIB superstring theory. In the previous section we introduced how to look for higher derivative corrections.
Given  the leading couplings in \reef{zzxc}, one needs to apply the higher derivative operators $\cD_{nm},\cD'_{nm}$ to \reef{zzxc} to be able to discover their all order corrections as follows
\beqa
&&{\cal L}^{n,m}= \pi^3\alpha'^{n+m+3}T_p\bigg(a_{n,m}\Tr\bigg[\cD_{nm}\left(\bPsi \ga^i D_b\Psi D^a\phi^i F_{ab} \right)+\cD_{nm}\left( \bPsi \ga^i D_b\Psi F_{ab} D^a\phi^i \right)
\nonumber\\&&+h.c \bigg]
+i b_{n,m}\Tr\bigg[\cD'_{nm}\left(\bPsi \ga^i D_b\Psi D^a\phi^i F_{ab} \right)+\cD'_{nm}\left(
 \bPsi \ga^i D_b\Psi F_{ab} D^a\phi^i  \right)+h.c.\bigg]
\bigg).
 \labell{Lnm681}
\eeqa
Now we need to extract the terms carrying the coefficients $a_{n,m}$ in \reef{Lnm681}
 to be able to derive the following vertex
\beqa
V_{\beta}^{b}(A,\bar\Psi,\Psi,\phi)=i\frac{T_p (2\pi\alpha')^3}{4} \bar u^A (\ga^j)_{AB} u^B \xi_{1j}\bigg( k_{1b}\frac{t}{2}\bigg)(t^n s^m+s^n t^m) a_{n,m}\Tr(\lam_1\lam_2\lam_3\lam_\beta).
\label{eesdde}\eeqa
Replacing \reef{eesdde} (instead of \reef{eesddem}) inside \reef{amp5211}  and also substituting the second term of the expansion of $L_3$ which appeared in (17)  inside \reef{ampc33} we might produce all infinite gauge poles of the amplitude. It is of high importance to mention the following remark as well. In order to consider $b_{n,m}$ coefficients, one needs to apply on-shell condition $(t+s+u=0)$ at each order of $\alpha'$ to the field theory vertices to obtain the desired terms in field theory amplitude.
\section{Conclusions}
In this paper we applied conformal field theory techniques and we found the complete form of the $<V_{C}V_{\bar\psi}V_{\psi} V_{\phi}>$ amplitude in IIB superstring theory. All infinite scalar/gauge (for $p+2=n,p=n$ cases) and fermion poles have been explored. We observed that the vertices of $V^{\alpha}_i(C_{p+1},\phi),V^{\alpha}_a(C_{p-1},A)$ do not require any corrections, hence,  all infinite  $(t+s+u)-$ channel scalar (gauge)  poles have provided worth information to determine infinite higher derivative corrections (to all orders in $\alpha'$) to two fermion-two scalar (two fermions-one scalar-one gauge) couplings which we have discovered them and particularly  their coefficients are exactly fixed.
  \vskip.02in
 We also clarified that the same universal conjecture for all higher derivative corrections that appeared in \cite{Hatefi:2012rx} holds for two fermion -two scalar couplings of IIB superstring theory.
\vskip.02in
It is worth pointing out that in RR vertex operator there are no winding modes so applying T-duality to the known results is not  effective. In particular in order not to miss any terms in superstring amplitudes and to be able to obtain all higher derivative corrections with their exact coefficients, one has to apply direct computations. Basically we proposed some patterns in this paper.
Let us talk about a subtle issue regarding the relation of open/closed string vertices in type II superstring theory.
For our amplitude which involves mixture open /closed strings, our calculations make sense of  using path integral formalism such that propagators (Green functions) are found by conformal field theory methods while the closed string has
$(\alpha_n,\tilde{\alpha}_n)$ oscillators. It is not obvious  how to do calculations with  oscillators, namely
in the first quantisation of strings it is subtle how to deal with $\tilde{\alpha}_n$.
Some comments have been suggested in (3.4) of \cite{Billo:2006jm} such that both oscillators for closed string would be
determined with open string's ones. In the other words some analytic continuation is needed and this kind of realisation implies that the state for closed string should be  considered
as a composite state of the open strings. The interpretation in field theory might be useful mentioning. It reveals that all background fields in DBI action should be some functions of super Yang-Mills fields. In the other words background fields must become composite  and  these functions would be examined once we employ
the complete open string formalism. Note also that supergravity background fields must include Taylor expansion as was argued in   \cite{Myers:1999ps}.
\section*{Acknowledgments}
I would like to thank J.Polchinski, J.Maldacena, R.Myers, W.Lerche, S.Stieberger, G.Veneziano, K.S.Narain, N.Lambert, A.Sagnotti and L.Alvarez-Gaume for valuable discussions.


\begin{thebibliography}{2007}
\bibitem{Polchinski:1995mt}
  J.~Polchinski,``Dirichlet-Branes and Ramond-Ramond Charges,''
  Phys.\ Rev.\ Lett.\  {\bf75}, 4724 (1995)
  [arXiv:hep-th/9510017].
\bibitem{Witten:1995im}
  E.~Witten,``Bound states of strings and p-branes,''
  Nucl.\ Phys.\  B {\bf 460},335 (1996)
  [arXiv:hep-th/9510135].
\bibitem{Polchinski:1996na}
  J.~Polchinski,`` Lectures on D-branes,''
  [arXiv:hep-th/9611050]
  ;
  C.~P.~Bachas,
  ``Lectures on D-branes,''
  [arXiv:hep-th/9806199].

\bibitem{Ademollo:1974fc}
  M.~Ademollo, A.~D'Adda, R.~D'Auria, E.~Napolitano, P.~Di Vecchia, F.~Gliozzi and S.~Sciuto,
  ``Unified Dual Model for Interacting Open and Closed Strings,''
  Nucl.\ Phys.\ B {\bf 77}, 189 (1974).

\bibitem{Polchinski:1996nb}
  J.~Polchinski,
   ``String duality: A Colloquium,''
  Rev.\ Mod.\ Phys.\  {\bf 68}, 1245 (1996)
  [hep-th/9607050].

\bibitem{Hatefi:2012sy}
  E.~Hatefi, A.~J.~Nurmagambetov and I.~Y.~Park,
  ``$N^3$ entropy of $M5$ branes from dielectric effect,''
  Nucl.\ Phys.\ B {\bf 866}, 58 (2013)
  [arXiv:1204.2711 [hep-th]].


\bibitem{Hatefi:2012bp}
  E.~Hatefi, A.~J.~Nurmagambetov and I.~Y.~Park,
   ``ADM reduction of IIB on $\mathcal{H}^{p,q}$ to dS braneworld,''
  JHEP {\bf 1304}, 170 (2013)
  [arXiv:1210.3825 [hep-th]].










\bibitem{Myers:1999ps}
  R.~C.~Myers,``Dielectric-branes,''
  JHEP {\bf 9912}, 022 (1999)
  [arXiv:hep-th/9910053].
\bibitem{Howe:2006rv}
  P.~S.~Howe, U.~Lindstrom and L.~Wulff,
   `On the covariance of the Dirac-Born-Infeld-Myers action,''
  JHEP {\bf 0702}, 070 (2007)
  [hep-th/0607156].
\bibitem{Leigh:1989jq}
  R.~G.~Leigh,
  ``Dirac-Born-Infeld Action from Dirichlet Sigma Model,''
  Mod.\ Phys.\ Lett.\ A {\bf 4}, 2767 (1989).
\bibitem{Cederwall:1996pv}
  M.~Cederwall, A.~von Gussich, B.~E.~W.~Nilsson and A.~Westerberg,
  ``The Dirichlet super three-brane in ten-dimensional type IIB supergravity,''
  Nucl.\ Phys.\ B {\bf 490}, 163 (1997)
  [hep-th/9610148]
  ;
  M.~Aganagic, C.~Popescu and J.~H.~Schwarz,
  ``D-brane actions with local kappa symmetry,''
  Phys.\ Lett.\ B {\bf 393}, 311 (1997)
  [hep-th/9610249]
;
  M.~Aganagic, C.~Popescu and J.~H.~Schwarz,
  ``Gauge invariant and gauge fixed D-brane actions,''
  Nucl.\ Phys.\ B {\bf 495}, 99 (1997)
  [hep-th/9612080]
  ;
  M.~Cederwall, A.~von Gussich, B.~E.~W.~Nilsson, P.~Sundell and A.~Westerberg,
  Nucl.\ Phys.\ B {\bf 490}, 179 (1997)
  [hep-th/9611159]
  ;
  E.~Bergshoeff and P.~K.~Townsend,
  Nucl.\ Phys.\ B {\bf 490} (1997) 145
  [hep-th/9611173].
\bibitem{Hatefi:2010ik}
  E.~Hatefi,
  ``On effective actions of BPS branes and their higher derivative
  corrections,''
  JHEP {\bf 1005}, 080 (2010)
  [arXiv:1003.0314 [hep-th]].

\bibitem{Hatefi:2012zh}
  E.~Hatefi,
   ``Shedding light on new Wess-Zumino couplings with their corrections to all orders in alpha-prime,''
  JHEP {\bf 1304}, 070 (2013)
  [arXiv:1211.2413 [hep-th]].

\bibitem{Hatefi:2012wj}
  E.~Hatefi,
   ``On higher derivative corrections to Wess-Zumino and Tachyonic actions in type II super string theory,''
  Phys.\ Rev.\ D {\bf 86}, 046003 (2012)
  [arXiv:1203.1329 [hep-th]].

\bibitem{Park:2007mc}
I. Y. Park, ``Scattering on D3-branes,''
Phys. Lett. {\bf B660}, 583 (2008) [arXiv:0708.3452[hep-th]]
;
I.~Y.~Park,``Open string engineering of D-brane geometry,''
  JHEP {\bf 0808}, 026 (2008)
  [arXiv:0806.3330[hep-th]].
\bibitem{Koerber:2002zb}
  P.~Koerber and A.~Sevrin,
  ``The NonAbelian D-brane effective action through order alpha-prime**4,''
  JHEP {\bf 0210}, 046 (2002)
  [hep-th/0208044].

\bibitem{Keurentjes:2004tu}
  A.~Keurentjes, P.~Koerber, S.~Nevens, A.~Sevrin and A.~Wijns,
  ``Towards an effective action for D-branes,''
  Fortsch.\ Phys.\  {\bf 53}, 599 (2005)
  [hep-th/0412271].

\bibitem{Denef:2000rj}
  F.~Denef, A.~Sevrin and J.~Troost,
  ``NonAbelian Born-Infeld versus string theory,''
  Nucl.\ Phys.\ B {\bf 581}, 135 (2000)
  [hep-th/0002180].
\bibitem{Hashimoto:1996bf}
A.~Hashimoto and I.~R.~Klebanov,
``Scattering of strings from D-branes,''
Nucl.\ Phys.\ Proc.\ Suppl.\  {\bf 55B}, 118 (1997)
[arXiv:hep-th/9611214]
;
I.\ R.\ Klebanov and L.\ Thorlacius,
``The Size of p-branes,''
Phys. Lett.~ {\bf B371}, 51 (1996)[arXiv:hep-th/9510200]
 ;
S.S.\ Gubser, A.\ Hashimoto, I.R.\ Klebanov, and J.M.\ Maldacena,
``Gravitational lensing by $p$-branes,''Nucl.\ Phys. {\bf  B472}, 231 (1996) [arXiv:hep-th/9601057]
;
C.\ Bachas, ``D-Brane Dynamics,'' Phys. Lett.~ {\bf B374}, 37 (1996)[arXiv:hep-th/9511043]
;
  J.~Polchinski,``String duality: A colloquium,''
  Rev.\ Mod.\ Phys.\  {\bf 68}, 1245 (1996) [arXiv:hep-th/9607050];
  W.~Taylor,``Lectures on D-branes, gauge theory and M(atrices),''
  [arXiv:hep-th/9801182];
  C.~Vafa,``Lectures on strings and dualities,''
  [arXiv:hep-th/9702201]
;
  ;%
  M.~Billo, M.~Frau, F.~Lonegro and A.~Lerda,``N = 1/2 quiver gauge theories from open strings with R-R fluxes,''
  JHEP {\bf 0505},047 (2005)
  [arXiv:hep-th/0502084]
  ;
  M.~Billo, P.~Di Vecchia, M.~Frau, A.~Lerda, I.~Pesando, R.~Russo and S.~Sciuto,``Microscopic string analysis of the D0-D8 brane system and dual R-R
  states,''
  Nucl.\ Phys.\  B {\bf 526},199 (1998)
  [arXiv:hep-th/9802088].
  ;
  E.~Hatefi,
  ``Three Point Tree Level Amplitude in Superstring Theory,''
  Nucl.\ Phys.\ Proc.\ Suppl.\  {\bf 216}, 234 (2011)
  [arXiv:1102.5042 [hep-th]];
  S.~de Alwis, R.~Gupta, E.~Hatefi and F.~Quevedo,
  ``Stability, Tunneling and Flux Changing de Sitter Transitions in the Large Volume String Scenario,''
  JHEP {\bf 1311}, 179 (2013)
  [arXiv:1308.1222 [hep-th];
  E.~Hatefi,
  ``SuperYang-Mills, Chern-Simons couplings and their all order $\alpha'$ corrections in IIB superstring theory,''
  arXiv:1310.8308 [hep-th].

 \bibitem{Hatefi:2012wz}
  E.~Hatefi, A.~J.~Nurmagambetov and I.~Y.~Park,
  ``Near-Extremal Black-Branes with n*3 Entropy Growth,''
  Int.\ J.\ Mod.\ Phys.\ A {\bf 27}, 1250182 (2012)
  [arXiv:1204.6303 [hep-th]]

\bibitem{McOrist:2012yc}
  J.~McOrist and S.~Sethi,
   ``M-theory and Type IIA Flux Compactifications,''
  JHEP {\bf 1212}, 122 (2012)
  [arXiv:1208.0261 [hep-th]].




\bibitem{Hatefi:2012ve}
  E.~Hatefi and I.~Y.~Park,
  ``More on closed string induced higher derivative interactions on D-branes,''
  Phys.\ Rev.\ D {\bf 85}, 125039 (2012)
  [arXiv:1203.5553 [hep-th]].

\bibitem{Hatefi:2012rx}
  E.~Hatefi and I.~Y.~Park,
  ``Universality in all-order $\alpha'$ corrections to BPS/non-BPS brane world volume theories,''
  Nucl.\ Phys.\ B {\bf 864} (2012) 640
  [arXiv:1205.5079 [hep-th]].
%
\bibitem{Hatefi:2012cp} 
  E.~Hatefi,
  ``On D-brane anti D-brane effective actions and their corrections to all orders in alpha-prime,''
  JCAP {\bf 1309}, 011 (2013)
  [arXiv:1211.5538,[hep-th]].

\bibitem{Garousi:2007fk}
  M.~R.~Garousi and E.~Hatefi,``On Wess-Zumino terms of Brane-Antibrane systems,''
  Nucl.\ Phys.\  B {\bf 800}, 502 (2008)
  [arXiv:0710.5875 [hep-th]].
\bibitem{Hatefi:2008ab}
  M.~R.~Garousi and E.~Hatefi,``More on WZ actions of non-BPS branes,''
  JHEP {\bf 0903}, 08 (2009)
  [arXiv:0812.4216 [hep-th]].
\bibitem{Vafa:1996xn}
  C.~Vafa,
   ``Evidence for F theory,''
  Nucl.\ Phys.\ B {\bf 469}, 403 (1996)
  [hep-th/9602022].

\bibitem{Stieberger:2009hq}
  S.~Stieberger,`` Open, Closed vs.Pure Open String Disk Amplitudes,''
  [arXiv:0907.2211 [hep-th]].
\bibitem{Kennedy:1999nn}
  C.~Kennedy and A.~Wilkins,``Ramond-Ramond couplings on brane-antibrane systems,''
  Phys.\ Lett.\  B {\bf 464}, 206 (1999)
  [arXiv:hep-th/9905195].

 \bibitem{Chandia:2003sh}
  L.~A.~Barreiro and R.~Medina,``5-field terms in the open superstring effective action,''
  JHEP {\bf 0503},055 (2005)
  [arXiv:hep-th/0503182].
;
  R.~Medina, F.~T.~Brandt and F.~R.~Machado,``The open superstring 5-point amplitude revisited,''
  JHEP {\bf 0207},071 (2002)
  [arXiv:hep-th/0208121]
  ;
  L.~A.~Barreiro and R.~Medina
  ,JHEP {\bf 1210}, 108 (2012)
  [arXiv:1208.6066 [hep-th]]
  ;
  A.~Bilal,
  ,Nucl.\ Phys.\ B {\bf 618}, 21 (2001)
  [hep-th/0106062].




\bibitem{Polchinski:1998aa}
J.~Polchinski,``String theory '',Vol 2,Cambridge University Press,1998




\bibitem{Liu:2001qa}
  H.~Liu and J.~Michelson,
  ``*-trek 3: The Search for Ramond-Ramond couplings,''
  Nucl.\ Phys.\ B {\bf 614}, 330 (2001)
  [hep-th/0107172].


\bibitem{Kostelecky:1986xg}
  V.~A.~Kostelecky, O.~Lechtenfeld, W.~Lerche, S.~Samuel and S.~Watamura,
  ``Conformal Techniques, Bosonization and Tree Level String Amplitudes,''
  Nucl.\ Phys.\ B {\bf 288}, 173 (1987).
\bibitem{Friedan:1985ge}
  D.~Friedan, E.~J.~Martinec and S.~H.~Shenker,
  ``Conformal Invariance, Supersymmetry and String Theory,''
  Nucl.\ Phys.\ B {\bf 271}, 93 (1986).

\bibitem{Hartl:2010ks}
  D.~Haertl and O.~Schlotterer,
  ``Higher Loop Spin Field Correlators in Various Dimensions,''
  Nucl.\ Phys.\ B {\bf 849} (2011) 364
  [arXiv:1011.1249 [hep-th]].
\bibitem{Fotopoulos:2001pt}
  A.~Fotopoulos,``On (alpha')**2 corrections to the D-brane action for non-geodesic
  world-volume embeddings,''
  JHEP {\bf 0109}, 005 (2001)
  [arXiv:hep-th/0104146].

\bibitem{Hashimoto:1996kf}
  A.~Hashimoto and I.~R.~Klebanov,
   ``Decay of excited D-branes,''
  Phys.\ Lett.\ B {\bf 381}, 437 (1996)
  [hep-th/9604065].
\bibitem{Hatefi:2013mwa} 
  E.~Hatefi,
  ``All order $\alpha'$ higher derivative corrections to non-BPS branes of type IIB Super string theory,''
  JHEP {\bf 1307}, 002 (2013)
  [arXiv:1304.3711 [hep-th]];
  E.~Hatefi,
``Selection Rules and RR Couplings on Non-BPS Branes,''
  JHEP {\bf 1311}, 204 (2013)
  [arXiv:1307.3520].

\bibitem{Billo:2006jm}
  M.~Billo, M.~Frau, F.~Fucito and A.~Lerda,
  ``Instanton calculus in R-R background and the topological string,''
  JHEP {\bf 0611}, 012 (2006)
  [hep-th/0606013].

















%






\end{thebibliography}
\end{document}